\documentclass[graybox, envcountchap]{svmult}

\usepackage{mathptmx}        
\usepackage{amsmath}
\usepackage{amssymb}
\usepackage{color}
\usepackage{helvet}          
\usepackage{courier}         
\usepackage{dirtree}

\usepackage{makeidx}        
\usepackage{graphicx}        
\usepackage{subfig}

\usepackage{multicol}        
\usepackage[bottom]{footmisc}

\usepackage{hyperref}        
\hypersetup{colorlinks=true,urlcolor=blue}
\usepackage{comment}
\usepackage{aas_macros}

\usepackage[misc]{ifsym}

\makeindex             

\newcommand{\ion}[2]{#1\,\textsc{#2}}

\begin{document}


\title{Observations of Repeating Tidal Disruption Events}
\titlerunning{Repeating TDEs} 
\author{Iair Arcavi and Thomas Wevers}
\institute{Iair Arcavi (\Letter) \at Tel Aviv University, Tel Aviv, Israel, \email{arcavi@gmail.com}
\and Thomas Wevers \at Astrophysics \& Space Center, Schmidt Sciences, New York, NY 10011, USA }

\maketitle

\abstract{The disruption of stars by supermassive black holes (SMBHs) as tidal disruption events (TDEs) can produce transient flares across the electromagnetic spectrum. Such flares can be used to probe SMBH properties, accretion physics and stellar dynamics in galactic nuclei. However, it can be difficult to distinguish TDE flares and those arising from other accretion phenomena around SMBHs, such as active galactic nuclei disk instabilities. Repeated flares from the same location can be produced by either such instabilities or by repeated partial disruptions of the same star, with the former expected to be at random intervals and the latter roughly periodic. Repeating nuclear transients can thus be used to distinguish between full and partial TDEs, as well as between TDEs and other accretion phenomena. Here we review observations of such repeating nuclear flares at various wavelengths. As time domain surveys become wider, deeper and faster, and as they accumulate longer baselines of data, more repeating phenomena at various time scales are being revealed. While these events show the potential for constraining several open questions regarding SMBH accretion phenomena, the small number of cases so far, the limited observational coverage for many of them, and the overall diversity of emission and repetition properties observed make it difficult to robustly associate most cases with physical scenarios at this stage. However, in the near future, as more data accumulate, the nature of various classes of repeating nuclear transients should become more clear, and with it the nature of many of the one-time flares as well.}


\section{Introduction}

A tidal disruption event (TDE) occurs when a star passes close enough to a supermassive black hole (SMBH) to be torn apart by tidal forces \cite{Rees1988}. The determination of ``close enough'' is often made in relation to the tidal radius $R_T=R_*\left({M_{BH}}/{M_*}\right)^{1/3}$ (where $M_*$ and $R_*$ are the mass and radius of the star, respectively, and $M_{BH}$ is the mass of the SMBH\footnote{In principle, TDEs can occur also around stellar-mass black holes, however the tidal radius there is much smaller, and hence the chances of such an event are much lower.}). A star passing within this radius from the SMBH is expected to be fully disrupted. For stars arriving on nearly parabolic orbits, half of the stellar material will be ejected and half accreted onto the SMBH, possibly producing an observable flare. As such, a TDE is considered a terminal event for the star involved. 

However, the tidal radius above assumes a uniform density star. Considering that stars have dense cores surrounded by lower density envelopes, it is possible for only an outer portion of a star to be stripped by the SMBH following an encounter close to, but outside of, the tidal radius (with the exact distance and amount of stripping depending on the density profile of the star). In such ``partial TDEs'', the surviving star could return for more encounters, resulting in repeated TDE flares around the same black hole, spaced roughly periodically in time. Repeating TDEs are thus an important observational characteristic for identifying partial disruptions. Contrasting such events with full disruptions can help us better understand TDE emission mechanisms, correctly account for TDE rates, and even learn about stellar dynamics in distant galaxy nuclei. It could also characterize flares not related to TDEs, but rather to enhanced accretion of gas from a pre-existing active galactic nucleus (AGN) disk, which could repeat at random intervals. As such, flares from galactic nuclei (so-called ``nuclear flares'') which are seen to repeat could help discern between extreme AGN activity, partial TDEs and full TDEs. Regardless of the exact source of accretion, repeating TDEs can also serve as probes of accretion physics as the limited amount of material transferred in each stripping could result in an accelerated passage through accretion states.

In this chapter we define a repeating transient as having at least two distinct flares separated by an epoch of quiescence in which the flux returns to the level seen before the first flare. This criterion excludes events which repeat on timescales shorter than the flare duration. In such cases, it is not possible to distinguish between a repeating event and light curve structure or modulation of a single event. 


\section{Optical Events}

Several types of optical nuclear transients have been identified over the last decade, including those associated with TDEs. The first optical TDE to be discovered in real time was PS1-10jh \cite{Gezari2012}. \cite{Arcavi2014} found that it was part of a class of TDEs characterized by hot ($\sim10^4$\,K) long-lived (months to years) blackbody continua and broad ($\sim10^3$--$10^4$\,km\,s$^{-1}$) \ion{He}{ii} and/or H$\alpha$ emission lines in their optical spectra. Several tens of such events are now known (e.g. \cite{vanVelzen2020,Yao2023}), and we refer to them here simply as ``optical TDEs''. The emission properties of these events are not what was initially predicted for TDEs, which were expected to produce X-ray flares from their transient accretion disks. Nevertheless, optical TDE rates are found to decline sharply in galaxies with SMBHs heavier than $10^8\,M_{\odot}$ \cite{2018ApJ...852...72V, Yao2023}. This is the SMBH mass range for which the tidal radius of a Sun-like star is smaller than the event horizon of the SMBH, suppressing any observable flare. A sharp decline of the event rate around that threshold is thus strong evidence that these events are indeed TDEs.

A second class of events was identified by \cite{Trakhtenbrot2019}, following the discovery of a flare in the galaxy F01004-2237 by \cite{Tadhunter2017}. This class shows similar blackbody temperatures as those seen in optical TDEs, as well as H and \ion{He}{ii} emission, but with much narrower ($\sim10^2$--$10^3$\,km\,s$^{-1}$) lines. In addition, the events in this class show lines from the Bowen Fluorescence mechanism (identified, for example, by certain \ion{O}{iii} and \ion{N}{iii} transitions) \cite{Bowen1928}. The class has thus been dubbed Bowen Fluorescence Flares (BFFs). While their light curves rise on timescales similar to those of optical TDEs, they decline much more slowly over years rather than months. The discovery of some BFFs in previously-known AGN \cite{Makrygianni2023, Sniegowska2025} led to the suggestion that BFFs may be related to TDEs in AGN, or to instabilities in AGN accretion disks. The exact characteristics of BFF light curves and spectra have yet to be robustly defined and their rates yet to be constrained, as only a handful of events are known. Hereafter, events with emission lines and species similar to those seen in BFFs are classified as such. 

A third class of nuclear transients display high amplitude TDE-like flares but exhibit AGN-like spectra. Dubbed ``ambiguous nuclear transients'' or ANTs \cite{Neustadt2020,Holoien2022}, their nature remains unclear. 

Here we consider all three classes, denoting them as ``TDE'', ``BFF'' and ``AGN'' according to their spectral properties. This is not a statement about the nature of the flares in the BFF and AGN spectral classes as TDEs or not TDEs, but rather serves as an observational classification.

TDEs in general are expected to be intrinsically rare (relative to supernovae, for example) with rates between $10^{-5}$ to $10^{-4}$ events per galaxy per year \cite{Wang2004,Stone2016}. However, \cite{Arcavi2014} found that optical TDEs (including some candidate repeaters; see below) show a strong preference for post-starburst galaxies. Such galaxies show an enhancement of the TDE rate of up to two orders of magnitude \cite{French2016,French2020}. In galaxies with the most extreme Lick H$\delta_A$ index (which correlates with the enhancement rate), seeing two unrelated optical TDEs from the same SMBH within several years of each other has a probability of a few percent (assuming Poisson statistics). Thanks to wide-field time domain surveys, the optical sky is monitored extensively, offering repetitions of transients to be detected on years and even decade time scales. Given that several tens of optical TDEs have been observed, it is thus possible that at least one of the repeating flares described below occurring in post-starburst galaxies is actually two unrelated TDEs which happened at the lower boundary of intervals between events.

In this section, we compile optical events for which a potential repetition has been published, and which have spectroscopic observations of at least one of the flares. 
Our list of repeating candidate optical nuclear transients is given in Table \ref{tab:optical_sample}. The requirement above for an epoch of quiescence between flares removes from our sample the possible repeating optical nuclear transients AT\,2019avd \cite{Malyali2021,Chen2022}, AT\,2019ehz \cite{Yao2023,Zhong2025} AT\,2019fdr \cite{Reusch2022}, AT\,2020acka \cite{Soraisam2022,Yao2023,Zhong2025}, AT\,2021loi \cite{Makrygianni2023}, AT\,2021uqv \cite{Yao2023,Zhong2025}, AT\,2021uvz \cite{Langis2026}, and AT\,2022exr \cite{Langis2026}. 

\begin{table}[]
    \centering
    \caption{Repeating Optical TDE Candidates}
    \label{tab:optical_sample}
    \begin{tabular}{llllll}
        \hline\noalign{\smallskip}
        Name & Repeat    & Host & No. of & Spectral & References \\
             & Timescale &      & Flares & Class    &  \\
        \noalign{\smallskip}\hline\noalign{\smallskip}
        AT\,2019azh  & 13.2 years & Post-starburst & 2 & TDE & \cite{Yao2026} \\
        AT\,2020vdq  & 3 years & Post-starburst & 2 & TDE & \cite{2025ApJ...985..175S}\\
        AT\,2022dbl  & 2 years & Post-starburst & 2 & TDE & \cite{2024ApJ...971L..26L, 2024arXiv241215326H, 2025ApJ...987L..20M}\\
        AT\,2023adr & 1 year & Unknown & 2 & TDE & \cite{LlamasLanza2024,Quintin2025,Angus2026} \\   
        AT\,2024pvu  & 17.1 years & Unknown & 2 & TDE & \cite{Langis2026,Yao2026} \\
        \noalign{\smallskip}\hline\noalign{\smallskip}
        F01004-2237 & 10 years & AGN & 2 & BFF & \cite{2024A&A...692A.262S} \\
        AT\,2019aalc  & 4 years & AGN & 2 & BFF & \cite{Veres2026} \\
        AT\,2021aeuk  & 3 years & AGN & 2 & BFF & \cite{Bao2024,2025ApJ...982..150S}\\
        AT\,2023uqm  & 425 days & Quiescent$^1$ & 5 & BFF & \cite{Wang2025} \\
        \noalign{\smallskip}\svhline\noalign{\smallskip}
        ASASSN-14ko & 114 days & AGN & $\gtrsim$10 & AGN & \cite{2021ApJ...910..125P, Tucker2021, 2023ApJ...951..134P, 2023ApJ...956L..46H, Huang2025}\\
        AT\,2022sxl  & 7.2 years & AGN & 2 & AGN & \cite{Ji2025}\\
        \noalign{\smallskip}\hline\noalign{\smallskip}
    \end{tabular}\\
    $^1$ Possible weak AGN.
\end{table}

\subsection{Events with TDE Spectra}

\subsubsection{AT\,2019azh}

AT\,2019azh is one of the best observed optical TDEs. It was extensively studied by \cite{Hinkle2021} and \cite{Faris2024}, and was part of the compilation of \cite{Yao2023}. Given the dense and early coverage, subtle structure in the light curve was identified and the first characterization of early spectroscopic properties was made \cite{Faris2024}. AT\,2019azh was considered to be a ``normal'' single-flare optical TDE, until \cite{Yao2026} discovered a flare from 2005 at the same position using archival Catalina Real-time Sky Survey (CRTS) \cite{Drake2009} $V$-band data. Only the declining phase of the 2005 flare was detected, showing a decline time scale a factor of a few shorter than that of the 2019 flare \cite{Yao2026}. As an archival event, no spectra or multi-band photometry of the 2005 flare are available, and confirming its nature as a TDE is difficult. \cite{Dgany2023} showed that Type Ia supernovae are $\sim$40 times more common observationally than TDEs in galaxy centers and indeed, such a supernova occurred within 2 years of the TDE AT\,2021mhg at the same position \cite{2025ApJ...985..175S}. 
Even if the 2005 flare were a TDE, it could be from an unrelated star. The host galaxy of AT\,2019azh has one of the most extreme Lick H$\delta_A$ index values observed in optical TDEs \cite{Hinkle2021}, indicating a TDE rate that is enhanced by $\sim$2 orders of magnitude compared to the TDE rate in ``normal'' (i.e. non post-starburst) galaxies. 

\subsubsection{AT\,2020vdq}

AT\,2020vdq is another candidate repeating optical TDE discovered through an archival search \cite{2025ApJ...985..175S}. Spectra of the second flare classify it as an optical TDE, but unfortunately there are no published spectra of the first flare, which occurred 2.6 years earlier. Thus, the nature of the first flare, which was cooler and roughly an order of magnitude fainter than the second one, can not be securely determined. If the first flare of AT\,2020vdq were a TDE, it would be one of the faintest and lowest-temperature TDEs ever observed \cite{2025ApJ...985..175S}. Even then, it could be an unrelated disruption. \cite{2025ApJ...987L..20M} use the Lick H$\delta_A$ index of its host to show that the probability of observing two unrelated TDEs within 2.6 years of each other in such a host galaxy is 0.86--8.61\%, depending on the assumption of the global TDE rate ($10^{-5}$--$10^{-4}$ events per galaxy per year, respectively).

\subsubsection{AT\,2022dbl}

AT\,2022dbl was an unremarkable optical TDE, showing characteristic broad He II emission together with weak H$\alpha$ emission, situating it in between H-rich and H-poor optical TDEs. What was remarkable was that approximately two years after the first flare, a second flare, very similar photometrically and identical spectroscopically, occurred \cite{2024ApJ...971L..26L, 2024arXiv241215326H, 2025ApJ...987L..20M}. 

Photometrically, the second flare was roughly a factor of two times fainter than the first and showed a slower decline. \cite{2025ApJ...987L..20M} found that the first decline was consistent with a $t^{-9/4}$ slope predicted for partial TDEs \cite{Coughlin2019}, while the second decline was better described by a $t^{-5/3}$ slope characteristic of full TDEs \cite{Rees1988,Phinney1989}. Despite these differences, the total energy outputs of both flares were similar to each other and to those of other TDEs \cite{2025ApJ...987L..20M}. 

\begin{figure}
\sidecaption
\includegraphics[width=0.6\textwidth]{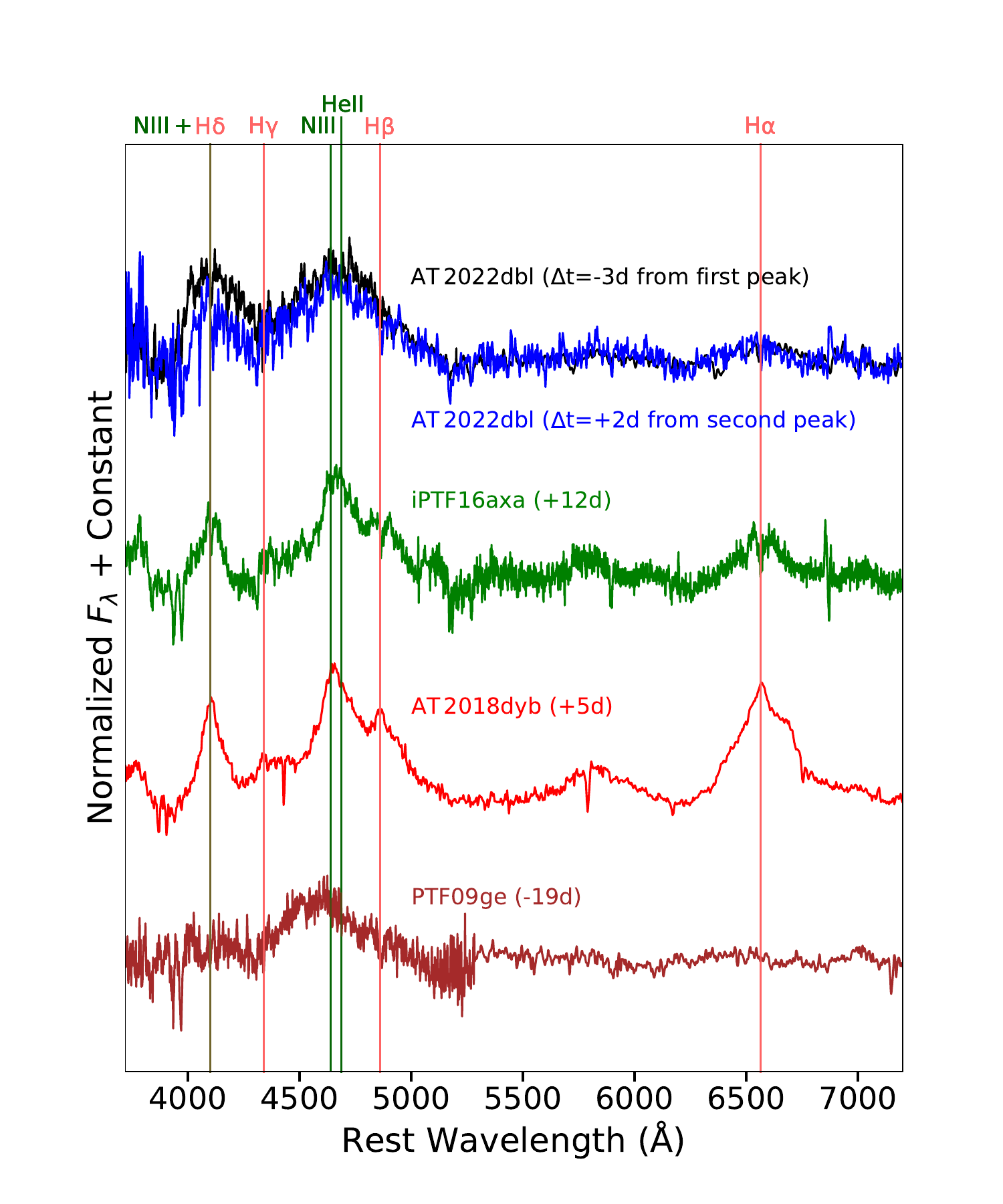}
\caption{Continuum-subtracted spectra of the first and second flare of AT\,2022dbl (top) showing nearly identical features, compared to the observed spread of H$\alpha$ strengths of events from the literature (spanning no H$\alpha$ emission at all to emission as strong as the broad \ion{He}{II} seen in all events shown here). All spectra were taken near peak luminosity (phases are noted in rest-frame days relative to peak luminosity). Figure from \cite{2025ApJ...987L..20M}.}
\label{fig:at2022dbl}      
\end{figure}

Spectroscopically, the two flares were virtually identical, showing the same level of weak H$\alpha$ emission (while the TDE population spans a continuum from no broad H emission to strong broad H emission; see Figure \ref{fig:at2022dbl}). Although the source of emission lines in optical TDEs is not known, \cite{2025ApJ...987L..20M} show that the strength of H$\alpha$ is not correlated with SMBH mass, and hence it might be tied to the properties of the disrupted star and/or its disruption orbit. Together, this was considered strong evidence by \cite{2025ApJ...987L..20M} that the two flares represent the repeated disruption of the same star, with the first flare due to a partial disruption, and the second possibly the full disruption of the star. Indeed, no third disruption has been observed so far, over 800 days since the second flare (which occurred roughly 700 days after the first one). 

AT\,2022dbl occurred in a post-starburst galaxy, but one with a moderate H$\delta_A$ Lick index. \cite{2025ApJ...987L..20M} estimate that the chance of two unrelated TDEs occurring within two years in such a galaxy is 0.037--0.368\% (depending on the assumption of the global TDE rate), further strengthening the likelihood that AT\,2022dbl indeed represents a repeating TDE, with at least the first encounter being a partial disruption. The fact that AT\,2022dbl was a ``normal'' optical TDE, with each flare showing typical spectral features, temperatures, luminosities and total emitted energies, raises the question of whether partial and full disruptions show very similar emission properties, and if the current sample of optical TDEs is in-fact a mixed bag of both types.

\subsubsection{AT\,2023adr}

\begin{figure}
\sidecaption
\includegraphics[trim={0 15cm 0 0},clip,width=0.6\textwidth]{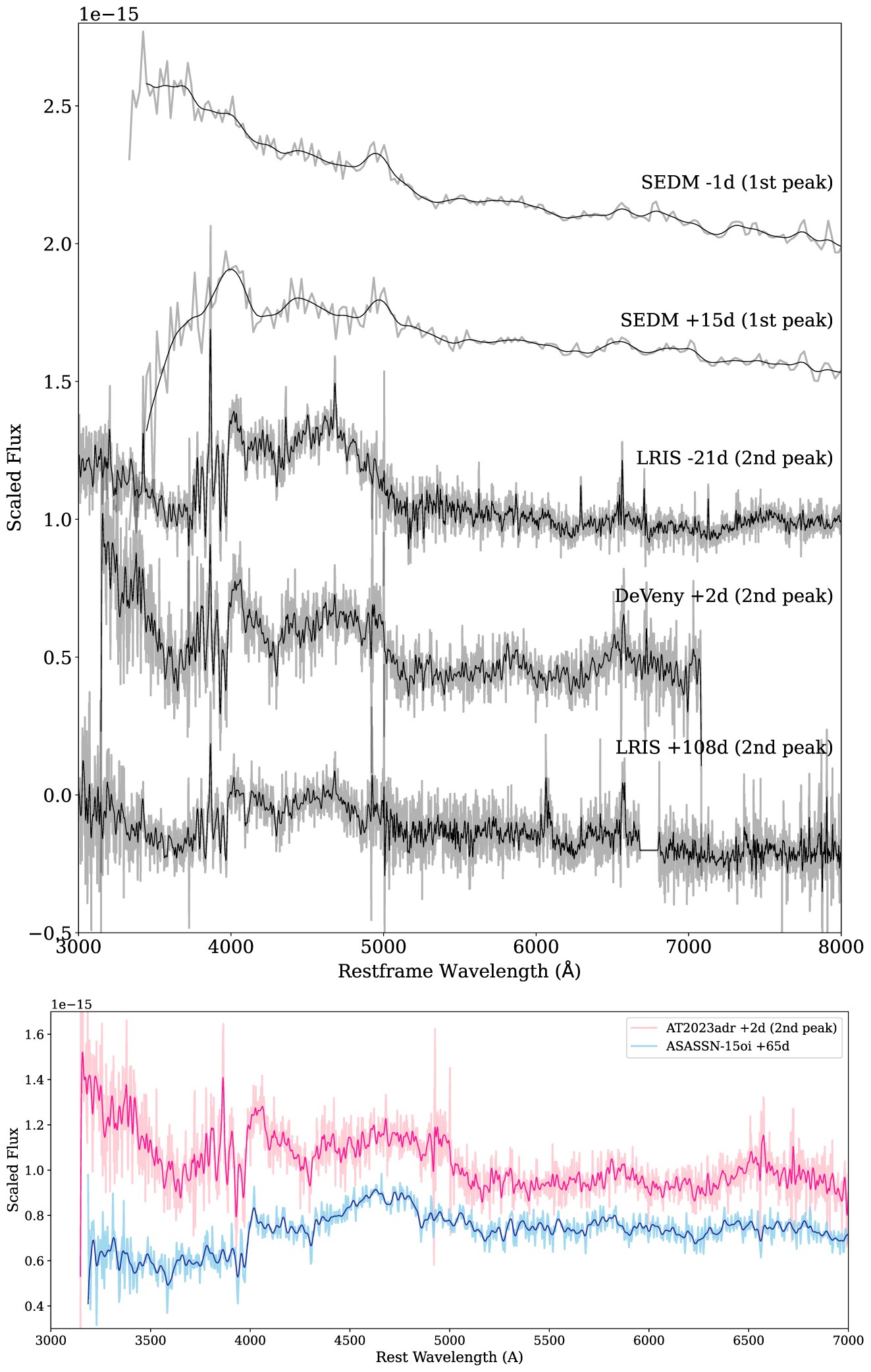}
\caption{Spectra of AT\,2023adr during the first peak show broad emission features consistent with H but not with \ion{He}{II}, while spectra during the second peak show typical TDE \ion{He}{II} emission. If these are two disrptions of the same star, the differing spectral features may hold imporant clues regarding the origin of spectral lines in TDEs in general - a current topic of investigation. Figure adapted from \cite{Angus2026}.}
\label{fig:at2023adr}      
\end{figure}

AT\,2023adr showed a persistently blue flare peaking at an absolute magnitude of roughly $-20$ to $-21$, coincident with the center of an otherwise quiescent galaxy \cite{Quintin2025}, all properties consistent with an optical-UV TDE. Low resolution spectra taken during the flare show possible broad features including H$\alpha$ and H$\beta$ with no clear signature of \ion{He}{II} \cite{Angus2026}. A year later, a second flare peaking approximately 1 magnitude fainter than the first flare, was discovered at the same location \cite{LlamasLanza2024,Quintin2025}. Only a single epoch of color information is available for the second flare, but a spectrum obtained then shows clear TDE signatures, including a blue continuum and broad H and \ion{He}{II} emission \cite{Angus2026} (Fig. \ref{fig:at2023adr}). No subsequent flares have been reported.

Both flares of AT\,2023adr are consistent with an optical-UV TDE, making AT\,2023adr a very likely repeating stellar disruption. However, the differing spectral features between the first and second flare differentiate this event from AT\,2022dbl. Spectra obtained during the second flare \cite{Shlentsova2024,Angus2026} show narrow emission lines of H and \ion{O}{III}, likely associated with the host galaxy. This suggest that the host is not a post-starburst galaxy, as is the case for many TDEs. Given the low rate of TDEs in non post-starbrust galaxies, a scenario of two unrelated TDEs is therefore highly unlikely. An unrelated superluminous supernova was initially suspected given the relatively long rise time and luminous peak magnitude \cite{Perley2023}, but the persistent blue colors make this scenario less likely. A clean spectrum of the host galaxy (not taken during a flare) would help determine whether it is a starforming galaxy or not. The latter option would rule out a superluminous supernova scenario and solidify the repeating TDE interpretation. In that case, having two flares with different spectral features could provide important clues regarding the not well-understood origin of TDE spectral lines.

\subsubsection{AT\,2024pvu}

AT\,2024pvu was classified as an optical TDE spectroscopically by \cite{Stein2024}. An archival flare at the position of AT\,2024pvu occurring 17.1 years earlier was discovered by \cite{Langis2026} and \cite{Yao2026} in CRTS data. \cite{Yao2026} used a GALEX serendipitous detection of the archival flare in both the NUV and FUV bands near its peak, together with the CRTS data, to infer a blackbody temperature of order $10^4$\,K and a bolometric luminosity of order $10^{44}$\,erg\,s$^{-1}$, both consistent with that of optical TDEs. These values are also consistent with those derived for the second flare by \cite{Yao2026} using available ZTF, ATLAS and Swift UVOT data. 

\cite{Yao2026} infer a SMBH mass of order $10^{7.9}$\,$M_{\odot}$, which is near the upper mass limit for suppression of TDE flares. Therefore, they argue that two unrelated TDEs within 17 years in such a galaxy is an unlikely scenario, and consider AT\,2024pvu to be the repeated disruption of the same star. 

\subsection{Events with BFF Spectra}

\subsubsection{F01004-2237}

F01004-2237 (named after its host galaxy IRAS F01004-2237) was initially classified by \cite{Tadhunter2017} as an optical TDE given its blue continuum and strong H and \ion{He}{ii} emission features. \cite{Trakhtenbrot2019} later showed that the emission features in F01004-2237 were much narrower, and that the light curve showed a much slower decline, compared to typical optical TDEs. In addition, \cite{Trakhtenbrot2019} identified prominent Bowen Fluorescence emission lines in the spectra, making F01004-2237 the first BFF. 

\cite{2024A&A...692A.262S} report a second flare at the same position as F01004-2237 roughly 10 years after the first flare. The light curves and spectra of both flares show similar properties, which \cite{2024A&A...692A.262S} claim are consistent with originating in a TDE. Yet even within this interpretation, it remained unclear if both flares represent the disruption of the same star, of two stars in a binary system, or of two unrelated stars.

\subsubsection{AT\,2019aalc}

AT\,2019aalc is a BFF which showed a second flare approximately four years after the first one \cite{Veres2026}. Both optical flares have similar characteristics with the second flare approximately 50\% brighter than the first, but showing a similarly slow decline and multiple light curve bumps during the decline (such ``rebrightenings'' are seen in the decline of many BFFs \cite{Makrygianni2023}). \cite{Veres2026} were able to rule out the presence of similar flares in the 14 years prior to the first one. They suggest that any future flares, if occurring at regular intervals, could favor a repeating TDE scenario over extreme AGN activity (which would likely be stochastic rather than periodic). 

\subsubsection{AT\,2021aeuk}

The triple flares of the BFF AT\,2021aeuk were investigated by \cite{Bao2024} and \cite{2025ApJ...982..150S}. The main flare in 2021 was preceded by a weaker event one year earlier, and was followed by a strong flare approximately three years later. The 2021 flare shows several bumps during its long decline, characteristic of BFFs. Spectra taken during the 2021 flare by \cite{2025ApJ...982..150S} are also more similar to those of BFFs than to those of optical TDEs. Archival spectra classify the host galaxy as harboring an AGN, specifically a narrow-line Seyfert 1. This too is in line with hosts of other BFFs. \cite{Bao2024} thus do not determine whether AT\,2021aeuk is a repeating TDE or some form of enhanced AGN activity. \cite{2025ApJ...982..150S}, on the other hand, point out subtle differences between the spectra of AT\,2021aeuk and those of other BFFs, and prefer a repeating TDE origin for AT\,2021aeuk. Recently a new flare of AT\,2021aeuk was reported \cite{Wang2026}, and is ongoing at the time of writing.

\subsubsection{AT\,2023uqm}

\begin{figure}
\includegraphics[width=\textwidth]{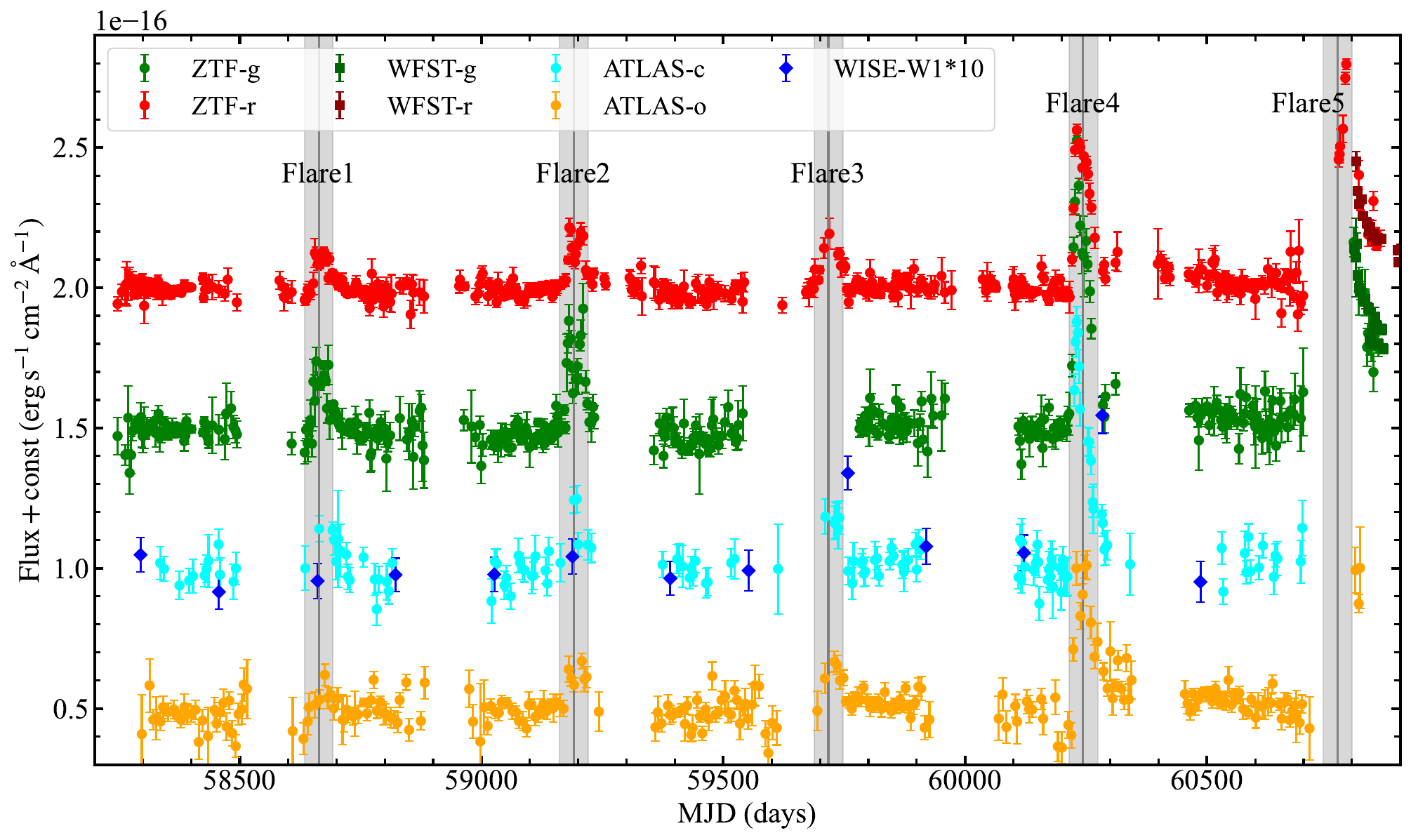}
\caption{Light curve of AT\,2023uqm. This is the only event other than ASASSN-14ko to show multiple periodic repetitions in optical wavelengths, and the only one to show progressively-increasing luminosity flares. Figure from \cite{Wang2025}.}
\label{fig:at2023uqm}       
\end{figure}

AT\,2023uqm showed five distinct flares at regular intervals of 526 observed days (corresponding to 425 restframe days) \cite{Wang2025}. The flares become progressively more luminous (Figure \ref{fig:at2023uqm}), and (for flares with enough coverage) show double peaks. The fifth (and most luminous) flare had a peak luminosity of order $10^{44}$\,erg\,s$^{-1}$ and blackbody temperature of $\sim$18,000\,K, similar to what is seen in optical TDEs. 

The spectra, however, taken during the fourth and fifth flares, are not similar to those of optical TDEs, but resemble more those of BFFs. They show relatively narrow Balmer lines as well as some Bowen Fluorescence lines (though the \ion{N}{iii} to \ion{O}{iii} line ratio is lower than that what is seen in some BFFs), \ion{Fe}{ii} lines, and high ionization coronal lines. 

The progressively increasing emitted energy, presumably from increasing mass being accreted across encounters, is connected by \cite{Wang2025} to partial disruption simulations that show that stars with convective envelopes experience runaway mass loss across encounters. Together with the measured \ion{N}{iii} to \ion{O}{iii} line ratio being lower than in BFFs, and similar to those of TDEs, \cite{Wang2025} interpret AT\,2023uqm as the progressive stripping of a low-mass ($\lesssim 1M_{\odot}$) star undergoing repeated partial TDEs. 

\subsection{Events with AGN Spectra}

\subsubsection{ASASSN-14ko}

\begin{figure}
\includegraphics[width=\textwidth]{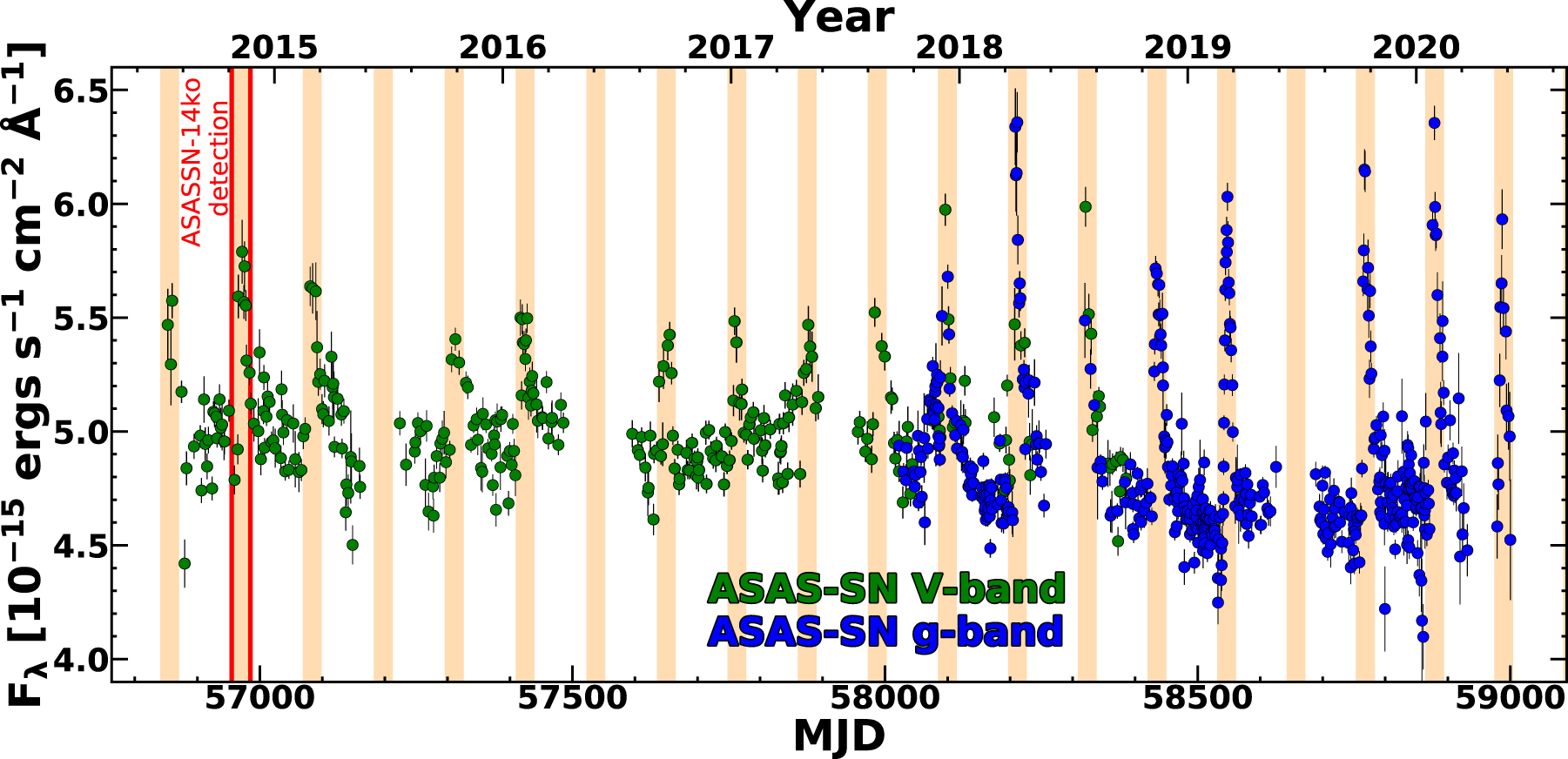}
\caption{Light curve of ASASSN-14ko. The shaded regions denote epochs of 30 days around model-predicted flare peaks. Figure from \cite{2021ApJ...910..125P}.}
\label{fig:asassn-14ko}       
\end{figure}

ASASSN-14ko is a repeating nuclear transient which displayed over 20 periodically spaced flares (Figure \ref{fig:asassn-14ko}) \cite{2021ApJ...910..125P, 2023ApJ...951..134P}. The period was seen to slowly decrease with time, having an average value of $115.2^{+1.3}_{-1.2}$ days and a rate of change of $\dot{P}=-0.0026\pm0.0006$ days per day \cite{2023ApJ...951..134P}. The flares are not identical, and are seen to change in UV luminosity and morphology over time \cite{2023ApJ...951..134P} with high-cadence observations of one flare revealing an early bump and late re-brigthening \cite{2023ApJ...956L..46H}. The optical output seems to be decreasing while the X-ray emission shows more sporadic behavior \cite{Huang2025}. UV spectra obtained during one of the flares shows blueshifted absorption lines that evolve within days (during the flare) to emission lines at their rest-frame wavelengths \cite{2023ApJ...951..134P}.

The blackbody radius, temperature and inferred bolometric luminosity of the optical emission in the flares was found to be similar to those seen in optical TDEs. However each flare evolved much more rapidly than typical optical TDEs, with a rise  of less than a week followed by an exponential decline. In addition, while in optical TDEs the temperature remains roughly constant and the radius declines, in the flares of ASASSN-14ko it seems that the radius remains roughly constant and the temperature declines (see Figure 9 in \cite{2021ApJ...910..125P}, though \cite{Huang2025} find a different behavior, see below). A further strong distinguishing feature between ASASSN-14ko and optical TDEs is its AGN-like optical spectrum seen during the flares. It shows narrow H$\alpha$, H$\beta$ and \ion{O}{iii} emission lines, with line ratios consistent with those seen in AGN (\cite{Tucker2021} find a second, unrelated AGN located $1.4\pm0.1$\,kpc from ASSASN-14ko). Nevertheless, a repeating partial TDE was considered the most favorable interpretation of ASASSN-14ko by \cite{2021ApJ...910..125P}. 

\cite{King2023} interpret ASASSN-14ko as a Quasi Periodic Eruption (QPE) source caused by repeated accretion from a white dwarf on an eccentric orbit, decaying to due gravitational wave emission, around a SMBH. Indeed, \cite{Huang2025} find that the blackbody temperature and radius in each outburst increase as the optical-UV luminosity increases, a trend also seen in QPEs. QPEs are typically atrributed to repeating X-ray sources, and are discussed further in Section \ref{sec:QPEs}.

\subsubsection{AT\,2022sxl}

AT\,2022sxl was initially classified as an AGN flare by \cite{Reguitti2022} given its spectrum which resembled those of narrow-line Seyfert 1 galaxies. Using archival CRTS and Gaia data, \cite{Ji2025} found evidence for a flare 7.2 years earlier at the same position. Both flares show similar luminosities and a similar multi-year decline. Together with the spectral features, this event is more consistent with a BFF or extreme AGN activity, than with an optical TDE.


\section{X-ray Events}

As mentioned above, the formation of a compact accretion disk should lead to bright X-ray emission. Indeed, a growing sample of TDEs and TDE candidate discoveries are being made in the X-ray band, including a number of potentially repeating events.

Table \ref{tab:xray_sample} summarizes the currently known sample of repeating X-ray nuclear transients not attributed to {\it normal} AGN variability. As in the optical case, with the increase in baselines of large-scale surveys, repeating variability has been detected on nearly all timescales, from hours to decades. 

When looking at the properties of these sources as a sample, some distinctions emerge when grouping sources by recurrence timescale rather than spectral type as done for the optical events. Below we summarize the prevalent properties for short, intermediate and long recurrence times of the X-ray repeating TDE candidates, with the caveat that given the small sample size, the full parameter space has not yet been fully explored. Future blind X-ray surveys as well as targeted X-ray follow-up observations of transients discovered at other wavelengths will likely continue adding to the diversity of observed behaviors.

\begin{table}[]
    \centering
    \caption{Repeating X-ray TDE Candidates}
    \label{tab:xray_sample}
    \begin{tabular}{llllll}
        \hline\noalign{\smallskip}
        Name & Repeat    & Host & No. of & Notes & References \\
             & Timescale &      & Flares &       &  \\
        \noalign{\smallskip}\svhline\noalign{\smallskip}
        Swift J0230 & 22 days & LLAGN & $>$10 & & \cite{2024NatAs...8..347G,2023NatAs...7.1368E, 2024arXiv241105948P}\\
        eRASSJ0456 & 223 days & Quiescent & 7 & & \cite{2023A&A...669A..75L, 2024A&A...683L..13L}\\
        XMMSL1J1404 & 710 days & Quiescent & 3 &  & \cite{2025arXiv251002905S} \\
        AT\,2018fyk & 1300 days & Quiescent & 2 & & \cite{2023ApJ...942L..33W,2024ApJ...971L..31P}\\
        GSN 069 & 9 years & Sy2 & 2 & QPEs & \cite{2023A&A...670A..93M}\\
        IC3599 & 9.5 years & LLAGN & 3 & & \cite{2015A&A...581A..17C}\\
        RXJ1331 & 30 years & Quiescent & 2 & & \cite{2023MNRAS.520.3549M} \\
        \noalign{\smallskip}\hline\noalign{\smallskip}
    \end{tabular}
\end{table}

\subsection{Timescales of Hours--Days: Quasi-Periodic Eruptions}\label{sec:QPEs}
QPEs are a distinct class of soft X-ray repeating nuclear events, characterized by high-amplitude, short-duration flares recurring on timescales of hours to 12 days (e.g. \cite{2019Natur.573..381M, 2020A&A...636L...2G}). The known sample consists of 13 sources, all discovered in the X-ray band (for a review, see the observational QPE chapter). Eruptions typically last between 2.5 hours and 3 days, reaching peak luminosities of 10$^{41}$--10$^{43}$ erg s$^{-1}$. Spectra are supersoft (kT $\approx$ 0.1 keV), with no significant emission detected above 2 keV. 

QPEs exhibit duty cycles generally between 10--30 per cent, and typical amplitudes of a factor 10--100 in soft X-ray flux. Some sources show alternating bright and faint flares (e.g. \cite{2019Natur.573..381M}) or switching between stable and more irregular phases \cite{2024A&A...692A..15G}. While most QPEs repeat steadily over months to years, others evolve or cease entirely on similar timescales. Their quiescent emission can be modeled as a compact accretion disk similar to those seen following TDEs \cite{2025ApJ...985..146G, 2025ApJ...980L...1W}; several sources have been discovered in late-time X-ray follow-up observations of optically and X-ray selected TDEs (e.g. \cite{2024Natur.634..804N, 2025ApJ...983L..39C, 2025MNRAS.540...30B}).

QPEs share many of the same host galaxy preferences as the broader TDE population, including low black hole masses, high stellar densities, centrally concentrated morphologies, and extended emission line regions \cite{2022A&A...659L...2W, 2024ApJ...970L..23W, 2025ApJ...994..209G}. The clean recurrence patterns and energetics of QPEs suggest a localized origin near the innermost regions of the accretion flow, potentially linked to disk instabilities or quasi-periodic perturbations from orbiting bodies. More specifically, \cite{Linial2023} suggest that QPEs are caused by the collisions between a main sequence star orbiting close to the SMBH as an extreme mass ratio inspiral (EMRI) and the expanding accretion disk produced by a recent TDE. The QPEs themselves are thus not considered to be repeating TDEs, and hence we do not list them in Table \ref{tab:xray_sample}, but QPEs offer a good reference point for comparison at the short timescale, high contrast end of repeating nuclear variability.

\subsection{Timescales of Weeks}

\subsubsection{Swift J0230}
\begin{figure*}
    \centering
    \includegraphics[width=0.9\linewidth]{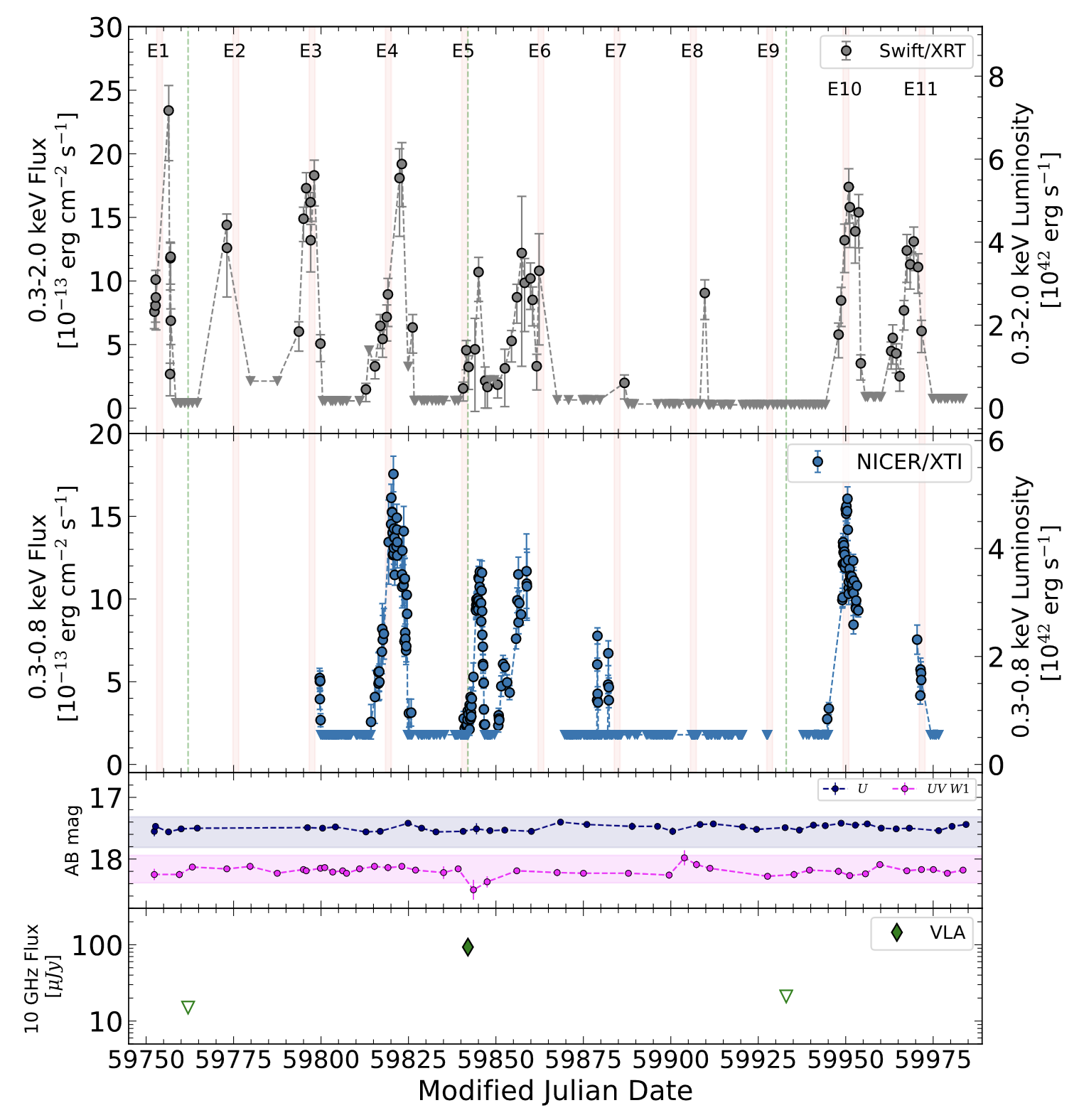}
    \caption{X-ray, UV/optical and radio lightcurves of Swift J0230, a nuclear transient repeating on $\sim$21-day timescales. No UV/optical variability was detected; a tentative radio source was detected during one observation. Figure from \cite{2024NatAs...8..347G}.}
    \label{fig:swj0230}
\end{figure*}

Swift J023017.0+283603 (Swift J0230+28) was an X-ray transient discovered by the Swift observatory in June 2022 \cite{2023NatAs...7.1368E, 2024NatAs...8..347G}. Over the course of $\sim$8 months, the source underwent a series of eight prominent, very soft X-ray eruptions roughly every $\sim$22–25 days (Fig. \ref{fig:swj0230}), making it one of the shortest recurrence timescale nuclear transients known outside QPEs. A Lomb–Scargle periodogram of the X-ray light curve shows a main peak around $\sim$22 days (with $\sim$1-day width). The outburst durations range between 10–15 days, although both shorter ($\sim$hours) and longer (30 days, possibly due to two overlapping bursts) eruptions are also observed \cite{2024arXiv241105948P}. Following this period of sudden activity, the eruptions then ceased and the source is now quiescent. 

Each eruption was characterized by a steep rise and fall in soft X-ray emission. The X-ray light curve was slightly asymmetric, with rises that were longer than the decays by $\sim$30\% on average. Rapid shut-offs (sometimes with a brief re-brightening or plateau just before) marked the end of each active phase. The peak X-ray luminosity was $\sim$4$\times$10$^{42}$ erg s$^{-1}$ (0.3–2 keV), making these outbursts as luminous as some TDEs or AGN flares. The quiescent X-ray flux is a factor of at least 100 lower than the flares \cite{2024NatAs...8..347G}.

The X-ray spectral properties of the eruptions were closer to those of QPEs than to those of longer timescale repeating nuclear transients. Namely, a very soft outburst spectrum, with no photons detected above 2 keV. A simple blackbody model (modified by Galactic absorption) describes the emission well, with a characteristic temperature kT $\sim$0.1 keV ($\sim$10$^6$ K). The temperature increases from $\sim$100 eV to 200 eV before decreasing again, although the highest temperature is reached after peak luminosity (i.e. during the decay phase). The lack of harder coronal X-ray emission is similar to QPEs and contrasts with sources classified as repeating TDEs, where a corona typically emerges at peak brightness if X-rays are observed. 

\subsection{Timescales of Months--Years} 

On longer timescales, the tendency for AGNs to exhibit large amplitude X-ray variability requires a more stringent set of observational constraints to unambiguously select repeating nuclear transients\footnote{In the optical, one result of this difficulty has been the adoption of a specific designation for transients whose properties do not allow a robust distinction between AGN outbursts and TDEs: ambiguous nuclear transients (ANTs; e.g. \cite{2020MNRAS.494.2538N, Hinkle2021}).}.
 
Four high fidelity repeating TDEs have been identified through the presence of high amplitude, repeating X-ray variability on (eRASSJ0456+20, AT\,2018fyk, XMMSL1J1404, and RXJ1331). These sources share several common properties that are indicative of similar physical processes at play. 

First, all four sources were found in quiescent host galaxies lacking both narrow and broad emission line regions, making an AGN explanation very unlikely (although it cannot be fully ruled out). 

Second, three of the sources show similar spectral variability in the form of soft to hard (thermally dominated to power-law dominated) X-ray spectral transitions (for the fourth, RXJ1331, data of sufficient quality are not available to determine whether it shows similar behavior). This is very unusual in the context of long-lived accretion episodes such a those in AGN. Even among (non-repeating) TDEs this is a relatively rare occurrence ($\sim$20 per cent of sources, \cite{2021MNRAS.508.3820S, 2024ApJ...966..160G, 2025A&A...697A.159G}). It is typically interpreted as the creation of the X-ray corona, in light of the similar evolution seen in X-ray binaries \cite{2021ApJ...912..151W}. 

A third common property appears to be the abrupt manner in which the X-ray emission disappears, decreasing by 2--3 orders of magnitude in brightness without an observationally identifiable precursor or indicator, rather than the more gradual power-law-like decay seen in the X-ray evolution of TDEs. 

These properties are broadly compatible with a scenario in which a star is captured through the Hills mechanism on an orbit with a period of $\sim$100s--1000 days, and which is repeatedly stripped of material following pericentre passage \cite{2023ApJ...942L..33W, 2023A&A...669A..75L}. Because they occur in quiescent galaxies, these events give rise to a nascent accretion flow, and the X-ray corona is initially absent; it can be observed to form in near real-time \cite{2021ApJ...912..151W} through high temporal cadence X-ray and UV monitoring. As the star returns to pericentre on a subsequent orbit, it disrupts the newly formed accretion flow. This leads to abrupt changes in the accretion process which manifest as sudden changes in X-ray/UV emission and/or the suppression of the harder coronal X-ray emission. 

We describe the evolution of the four known repeating nuclear X-ray transients in more detail below.

\subsubsection{eRASSJ0456}

\begin{figure*}
    \centering
    \includegraphics[width=\linewidth]{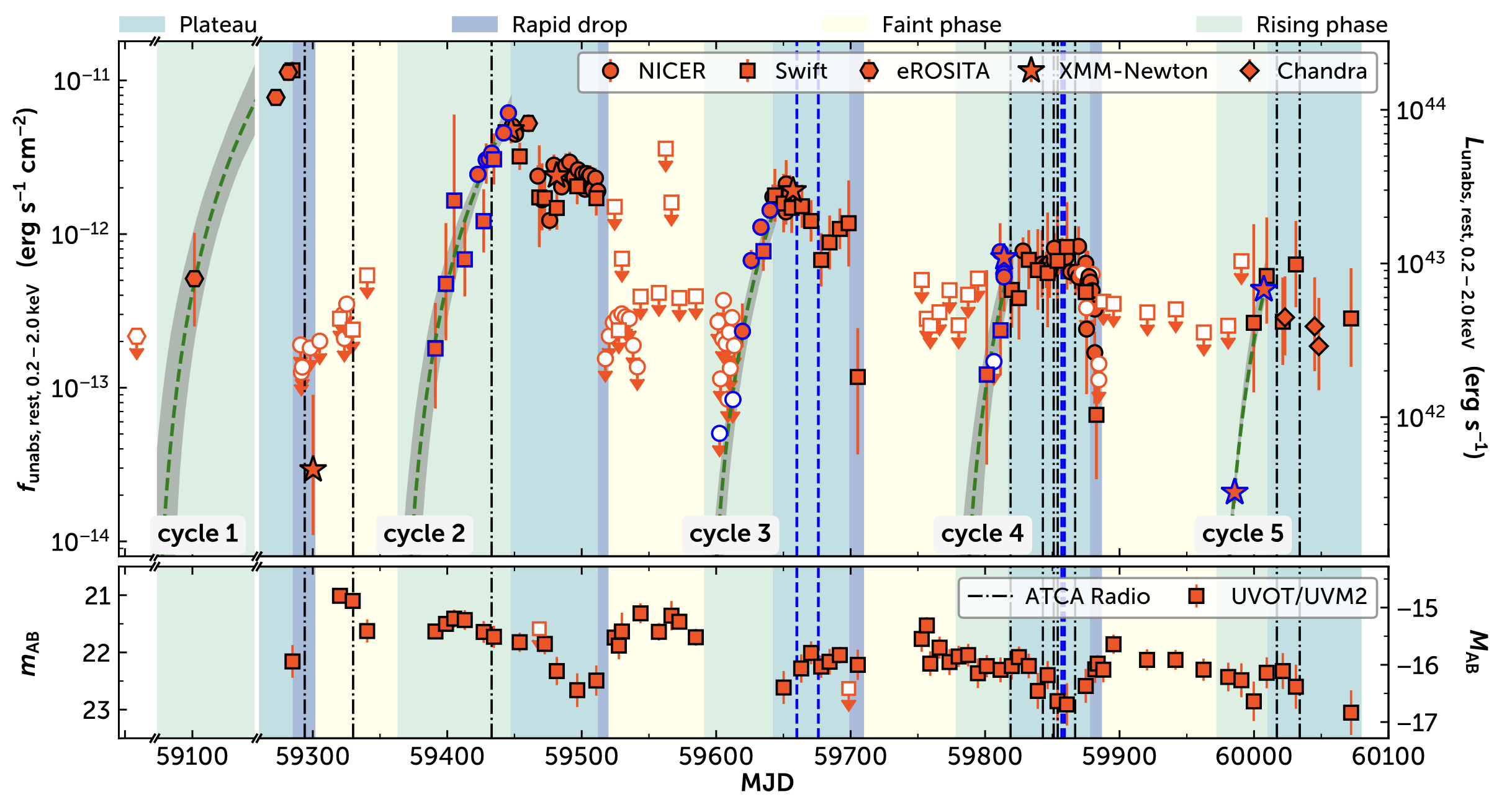}
    \caption{X-ray (top) lightcurves of eRASSJ0456--20, showing repeating flares that decrease in peak luminosity over time. No significantly correlated UV (bottom) variability is observed. Figure from \cite{2024A&A...683L..13L}.}
    \label{fig:erassj0456}
\end{figure*}

eRASSt J045650.3–203750 (eRASSJ0456–20) is a nuclear transient in a quiescent galaxy (z=0.077) that exhibits repeated X-ray flares \cite{2023A&A...669A..75L}. At least five distinct outbursts have been observed (Figure \ref{fig:erassj0456}), with an initial recurrence period of $\sim$300 days that decreased to $\sim$230 days and $\sim$190 days in subsequent cycles \cite{2024A&A...683L..13L}. In addition to the recurrence time, the duration of each of the active phases as well as the peak X-ray luminosity also decrease over time. Each cycle follows a distinct four-phase pattern: a slow X-ray rise (40--100 days) leading into a plateau phase (60--120 days), transitioning into a rapid drop in X-ray flux by over two orders of magnitude within $\sim$1 week, and a faint quiescent phase (60--80 days) before the next rise. The flares have a “slow-rise, fast-decay” morphology.  

The X-ray spectra evolve through each phase of the cycle. During the rise phase, the spectrum is ultra-soft (photon index $\Gamma >$ 3) with a thermal component that can be well described by a multi-color disk blackbody with a temperature kT $\sim$70 eV. As the flare evolves onto the plateau, the spectrum hardens, indicating the development of a coronal component: high-quality XMM-Newton data show evidence of a warm/hot corona contributing via inverse Comptonization of the disk’s soft seed photons. This suggests a transition from a disk-dominated soft state to a steep power-law state as the outburst progresses. Once the flare rapidly declines, the X-rays return to a faint soft state. 

In UV and optical wavelengths, eRASSJ0456–20 shows only low level variability. The UV flux does not track the X-ray flares, and no significant optical brightening is detected above the host galaxy light. This indicates that the outbursts are primarily an extreme-UV/X-ray phenomenon; the overall trend is a decrease of both the X-ray and UV luminosities over time. The transient was also detected in the radio band during some outbursts: a radio flare was recorded coincident with two X-ray plateaus, potentially indicating transient emission. However, the X-ray quiescent phase was not sufficiently sampled to confirm significantly correlated variability. 

\subsubsection{AT\,2018fyk}

Unlike the other repeating nuclear X-ray transients we have considered, AT\,2018fyk is the only source that was first discovered at optical wavelengths,
\cite{2019MNRAS.488.4816W}. Luminous X-ray emission of $>$10$^{44}$ erg s$^{-1}$ appeared shortly after the UV/optical peak. This is currently the only recurring X-ray transient that has been spectroscopically classified as an optical TDE, showing transient broad H and \ion{He}{ii} emission \cite{2019MNRAS.488.4816W}. 

Two outbursts of bright X-ray and UV emission have been observed so far, lasting for 500--600 days and with a repeat timescale of $\sim$1300 days \cite{2023ApJ...942L..33W, 2024ApJ...971L..31P}. There was no optical counterpart to the second outburst.
During the first flare, the X-ray spectrum evolved from a disk-dominated (thermal) state to a corona (non-thermal) dominated state, with the formation of the X-ray corona captured over $\sim$100s of days \cite{2021ApJ...912..151W}. The properties of the accretion flow (such as the evolution of the spectral energy distribution and the spectral state) during the second outburst are consistent with the properties of the first outburst \cite{2023ApJ...942L..33W}.
The end of the first flare was marked by a sharp drop in X-ray and UV luminosity, by factors of $>$6000 and 15 respectively, in $\sim$100 days. The second outburst also showed a significant drop, albeit more gradual than the first, by factors of $>$10 (X-ray) and $\sim$3 (UV) over $\sim$50 days.

\subsubsection{XMMSL1J1404}
XMMSL2 J1404-25 was discovered through the XMM Slew Survey in a quiescent galaxy \cite{2025arXiv251002905S}. The X-ray lightcurve shows short (days) timescale variability over a period of 100 days, before a steep ($\propto t^{-5.2}$) decline over 230 days, during which the X-ray emission faded by a factor of at least 500. Its spectral evolution is notable due to the rapid formation of a Comptonization zone, observed as the emergence of a hard spectral component, and attributed to the formation of the X-ray corona over a period of 7 days. No correlated UV variability was observed during the X-ray flare.

Two further rebrightening episodes were observed 2 and 4 years after the first flare. These repeated X-ray flares were discovered serendipitously in eROSITA scans, but higher cadence or sensitivity monitoring is not available to study their evolution in detail. Swift observations during the expected fourth flare revealed only non-detections, indicating that the flaring may have ended. 

\subsubsection{RXJ1331}

RX\,J133157.6$-$324319.7 (RXJ1331) was a repeating ultra-soft X-ray transient in a quiescent galaxy \cite{2023MNRAS.520.3549M}. The source was first detected in outburst in 1993 by \textit{ROSAT}, when it brightened by a factor of at least 40 relative to prior upper limits, reaching a 0.2--2\,keV flux of ${\sim}10^{-12}$\,erg\,s$^{-1}$\,cm$^{-2}$ , with an ultra-soft spectrum well described by a blackbody of $kT = 0.11 \pm 0.03$\,keV. The rise was rapid, with the flux increasing by a factor of eight over only ${\sim}8$\,days, and the source subsequently faded by a factor of at least 30 in pointed \textit{ROSAT} observations ${\sim}165$\,days later. 

After non-detections spanning nearly three decades in archival \textit{XMM} Slew, \textit{Swift} XRT, and the first four eROSITA All-Sky Surveys, RXJ1331 was re-detected in eRASS5 in January 2022 at a 0.2--2\,keV flux of $(6.0 \pm 0.7) \times 10^{-13}$\,erg\,s$^{-1}$\,cm$^{-2}$, with a spectrum ($kT = 0.115 \pm 0.007$\,keV) consistent with that of the 1993 flare \cite{2023MNRAS.520.3549M}. Pointed \textit{XMM} observations ${\sim}17$\,days after the eRASS5 detection constrained the flux to have decayed by a factor of ${\gtrsim}\,40$, indicating a similarly short decay timescale as the 1993 flare. 

Unlike the other long-timescale repeaters discussed previously, no soft-to-hard spectral transition has been observed in J1331, though the available data are of insufficient quality to firmly rule one out. The outbursts are not associated with any transient UV, optical, mid-infrared, or radio emission, with deep upper limits across all bands during follow-up. The extremely low Poisson probability (${\sim}5 \times 10^{-6}$) of observing two unrelated full TDEs from the same galaxy within 30 years, combined with the quiescent nature of the host and the short rise/decay timescales led \cite{2023MNRAS.520.3549M} to interpret RXJ1331 as repeated weak partial disruptions of a star on an elliptical orbit around the SMBH. Unfortunately, no high enough signal to noise ratio spectra of the host galaxy of RXJ1331 are currently available to determine its H$\delta_A$ Lick index in order to check if it could accommodate such an enhanced TDE rate.

\subsubsection{IC3599}

The first observational claim of a repeating TDE candidate was made for IC3599 by \cite{2015A&A...581A..17C}. IC3599 is a Seyfert 1.9 galaxy discovered as a very bright AGN in the ROentgen SATellite (ROSAT) \cite{Aschenbach1981} all-sky survey, gradually decreasing in brightness by a factor of $>$100. In addition to the dramatic X-ray flux changes, variability in the optical emission lines was observed, leading to the interpretation of the event as either a TDE or a changing-look AGN \cite{1995MNRAS.273L..47B, 1995A&A...299L...5G}. Fading coronal emission lines were also observed several years after the outburst \cite{1999A&A...343..775K}. After residing in a low flux state for $\sim$19 years\footnote{A tentative detection of a flare 9.5 years after the first relies on a single data point.}, Swift observed a second large amplitude outburst, leading \cite{2015A&A...581A..17C} to suggest the repeating TDE scenario. 

\cite{2015ApJ...803L..28G} instead argued that the long-term evolution and source properties were more likely to be associated with accretion disk instabilities of the AGN. {\bf More recent follow-up observations of the putative outburst expected in 2019-2020 did not yield a significant detection \cite{2024ApJ...969...98G}. This was initially interpreted as evidence that the variability was not produced by a repeating TDE. However, continued monitoring subsequently revealed a third giant outburst (with a luminosity change by a factor of $>$100), detected in real time several years after the originally predicted epoch \cite{2026ApJ..1006L..23G}. 

The outburst spectra are supersoft, with almost no photons detected above 2.5\,keV, and the source reached the Eddington limit at peak. Notably, and unlike the majority of the X-ray selected repeaters discussed in this chapter, the outburst was accompanied by flaring at UV and optical wavelengths, as well as a response of the optical emission lines, including the reappearance of high-ionization coronal lines \cite{2026ApJ..1006L..23G}. The \textit{XMM-Newton} light curve during the outburst also revealed an oscillatory QPO-like pattern. 

The delayed arrival of the third outburst relative to the predicted strict periodicity led \cite{2026ApJ..1006L..23G} to favor an accretion disk instability over a repeating TDE interpretation, although the debate first sparked by this source 30 years ago remains unsettled.

}

\subsubsection{GSN 069}
GSN 069 is an ultra-soft AGN hosting a low mass black hole ($\sim10^6$ M$_{\odot}$ showing repeated long term flaring \cite{2023A&A...670A..93M}; it is best known as the source in which QPEs were first identified \cite{2019Natur.573..381M}. The system was first detected in outburst in July 2010 during an \textit{XMM-Newton} slew \cite{2008A&A...480..611S}, at a flux level more than a factor of 240 above \textit{ROSAT} upper limits obtained 16 years earlier. The X-ray spectrum was extremely soft (kT $\sim 50$\,eV), with no detectable emission above ${\sim}1$\,keV, and the optical spectrum revealed only narrow emission lines, leading to a Seyfert 2-like classification of the nucleus.

The X-ray flux remained roughly constant for the first ${\sim}1$\,year \cite{2013MNRAS.433.1764M} before entering a smooth decay over the following ${\sim}7$--8 years \cite{2018ApJ...857L..16S}. 
The decay was interrupted in 2020, when the source rebrightened, an event interpreted by \cite{2023A&A...670A..93M} as a second TDE occurring ${\sim}9$ years after the first. Notably, a precursor X-ray flare was detected prior to the second outburst, possibly associated with the circularisation phase of disc formation, with a similar precursor tentatively identified before the first event \cite{2023A&A...670A..93M}. In contrast to the other long-timescale repeaters described above, GSN 069 has remained thermally dominated throughout its evolution: no soft-to-hard spectral transition or coronal component has ever been observed, and the decay of each outburst is gradual rather than ending in an abrupt shut-off. 

The repetition of two long-lived, ultra-soft accretion episodes from the same nucleus make GSN 069 a candidate for the repeated partial disruption of the same star. Its long-lived evolution is unusual compared to the rest of the sample, potentially due to differences in the viscous timescale of the accretion flow \cite{2025ApJ...992..114G}.


\section{Discussion}

While it is tempting to look for a unifying explanation (e.g. similar physical processes operating on different scales, leading to distinct observational properties), the large diversity and heterogeneity of the growing sample of repeating nuclear transients, even within each wavelength regime, seems to disfavor that scenario. 

In both the optical and X-ray wavelength regimes, a significant fraction of sources are hosted by nuclei classified as active based on classical diagnostic diagrams using narrow emission line ratios. 
While additional evidence is often cited to support classification as a (repeating) TDE, in the form of unusual X-ray spectral behaviour or the peculiar evolution of the spectral energy distribution, it is typically not possible to fully rule out alternative scenarios related to a pre-existing accretion flow. 

The source IC3599 exemplifies the typical community debate as to whether repeat variability, observed in X-rays in this case, is caused by a repeating TDE, or instead results from changes (intrinsic or observed) in a pre-existing accretion flow, as is typical behavior of AGNs. 
Similar debates still exist in the community 30 years later, also in the optical regime \cite{2021SSRv..217...54Z}. High pressure on scarce observational resources, especially given the increasing transient discovery rates, may prevent the acquisition of crucial follow-up data and lead to erroneous conclusions. In addition, even for TDEs in quiescent galaxies, some repeating TDEs might not be possible to distinguish from single TDEs around SMBH binaries (e.g. \cite{Wen2024}) or single TDEs with precessing disks (e.g. \cite{Chen2026}).

With a rapidly growing collection of astronomical data, mining archives for particular and unexpected modes of variability, such as the presence of multiple large amplitude flares, is becoming commonplace. Because even a single large amplitude flare is a rare occurrence in AGNs, statistical arguments are often used to argue that multiple such flares are inconsistent with normal AGN variability, and are more likely to be good repeating TDE candidates. 
This hides two misconceptions, likely related to an underestimate of the sample size that is being searched. 
First, the AGN flare amplitude distribution has a long tail towards high amplitudes. While \cite{Graham2017} find only 51 ``major flares'' (having a median peak amplitude of 1.25 magnitudes) out of 900,000 quasars, \cite{Rumbaugh2018} estimate that $\sim$30--50\% of quasars vary by more than 1 magnitude over a $\sim$15-year baseline. 
Second, typical AGN variability is governed by a red noise process. This has the unfortunate property (from the perspective of repeating TDE searches) of occasionally producing repeat flares that appear to be equidistantly spaced in time for 3--5 cycles, which is then interpreted as evidence for a (quasi-)periodic signal indicative of a repeating TDE. 

While the combination of equidistantly spaced, large amplitude flares have a very low statistical probability, dedicated searches over long baselines and large samples (10$^5$ or more sources) will inevitably yield events in the extreme tails of the distributions. Dedicated searches deliberately focus on these tails, and can misinterpret their distance to the distribution center as evidence of a distinct physical interpretation. 

Exacerbating this problem is the fact that such searches are typically performed on sparsely sampled lightcurves and/or as archival searches years after the flares occurred, making real-time follow-up to further investigate the nature of the variability impossible.

Indeed, in their systematic search of a volume-limited sample, \cite{Yao2026} conclude that repeating TDEs with intervals of $<$20 years may account for 25\%-–60\% of the optical TDE sample. This high rate has been referred to as ``a formation crisis for repeating TDEs'' \cite{Pan2026}, suggesting that at least some events identified as repeating TDEs may be of different origins.

The best candidates for genuine repeating TDEs (as opposed to stochastic AGN variability) are sources found in quiescent host galaxies, sources with optical spectral TDE classifications of more than one flare,  and/or sources where a large number of flares have been observed. Large amplitude flaring is very unusual in galaxies lacking other indicators of long-lived episodes of accretion in the form of broad or narrow emission line regions and/or archival X-ray detections. Similarly, the detection of many repeat outbursts allows for a detailed investigation of the similarities and differences between flares, and whether the properties of each flare remain consistent over time. This usually provides sufficient evidence to distinguish between AGN variability and repeating TDEs.

Of course, sampling multiple outbursts is typically only possible if the repeat timescales are short (10s--100s of days). There is thus likely an observational bias in the existing sample towards shorter recurrence times. While this is likely to remain the case for the X-ray selected samples, the densely sampled, large area sky coverage provided by optical surveys 
is reaching a sufficiently long baseline and high enough cadence that repeating transients with recurrence times of 1000s of days are starting to be detectable in growing numbers. In light of the pitfalls discussed in this Chapter, we encourage the community to adopt a high burden of proof to claim sources as repeating TDEs. 


\begin{acknowledgement}
I.A. acknowledges support from the European Research Council (ERC) under the European Union’s Horizon 2020 research and innovation program (grant agreement number 852097) and from the United States - Israel Binational Science Foundation (BSF; grant number 2024812).
\end{acknowledgement}

\bibliographystyle{spphys.bst}
\bibliography{bibliography.bib}


\end{document}